\documentclass[conference]{IEEEtran}
\IEEEoverridecommandlockouts
\usepackage{cite}
\usepackage{amsmath,amssymb,amsfonts}
\usepackage{algorithmic}
\usepackage{graphicx}
\usepackage{textcomp}
\usepackage[table,xcdraw]{xcolor}
\usepackage{fixltx2e}
\usepackage{subscript}
\usepackage{soul}
\usepackage{color}
\newcommand{\citet}{\cite}

 \usepackage[T1]{fontenc}
 \usepackage[utf8]{inputenc}

\usepackage[pdfencoding=auto]{hyperref}
\usepackage{bookmark}
\usepackage{multirow}

\usepackage{units}

\usepackage{soul}
\usepackage{color}

\usepackage{makecell}

\usepackage{enumitem}

\usepackage[table,xcdraw]{xcolor}
\usepackage{listings}
\usepackage[frozencache,cachedir=.,newfloat]{minted}
\usepackage{caption}
    
\newenvironment{code}{\captionsetup{type=listing}}{}
\SetupFloatingEnvironment{listing}{name=Snippet}
\definecolor{LightGray}{HTML}{444444}

\def\BibTeX{{\rm B\kern-.05em{\sc i\kern-.025em b}\kern-.08em
    T\kern-.1667em\lower.7ex\hbox{E}\kern-.125emX}}

\begin{document}

\title{Automated Artefact Relevancy Determination from Artefact Metadata and Associated Timeline Events}


\author{
    \IEEEauthorblockN{Xiaoyu Du\IEEEauthorrefmark{1}, Quan Le\IEEEauthorrefmark{2}, Mark Scanlon\IEEEauthorrefmark{3}} 
    \IEEEauthorblockA{\IEEEauthorrefmark{1}\IEEEauthorrefmark{3}Forensics and Security Research Group, University College Dublin, Ireland.
    \\xiaoyu.du@ucdconnect.ie, mark.scanlon@ucd.ie}
    \IEEEauthorblockA{\IEEEauthorrefmark{2}\IEEEauthorrefmark{3}Centre for Applied Data Analytics Research (CeADAR), University College Dublin, Ireland.
    \\quan.le@ucd.ie}
}

\maketitle

\begin{abstract}
Case-hindering, multi-year digital forensic evidence backlogs have become commonplace in law enforcement agencies throughout the world. This is due to an ever-growing number of cases requiring digital forensic investigation coupled with the growing volume of data to be processed per case. Leveraging previously processed digital forensic cases and their component artefact relevancy classifications can facilitate an opportunity for training automated artificial intelligence based evidence processing systems. These can significantly aid investigators in the discovery and prioritisation of evidence. This paper presents one approach for file artefact relevancy determination building on the growing trend towards a centralised, Digital Forensics as a Service (DFaaS) paradigm. This approach enables the use of previously encountered pertinent files to classify newly discovered files in an investigation. Trained models can aid in the detection of these files during the acquisition stage, i.e., during their upload to a DFaaS system. The technique generates a relevancy score for file similarity using each artefact's filesystem metadata and associated timeline events. The approach presented is validated against three experimental usage scenarios.
\end{abstract}


\section{Introduction}
\label{intro}

The ``Golden Age" in digital forensics is the period from the 1990s to the first decade of the twenty-first century~\cite{garfinkel2010digital}. Since then, the diversity of personal digital devices, the vast amount of data storage, and the prevalence of cloud services each bring compounding challenges to digital forensic investigations~\cite{scanlon2016battling}. The vast amount of data being encountered by law enforcement agencies throughout the world can not be analysed in a timely manner. This has lead to significant multi-year digital forensic backlogs becoming commonplace~\cite{lillis2016current}. The increasing number of cases requiring digital forensic investigation, coupled with their associated large data volumes, are difficult to process using existing investigation techniques~\cite{mohammed2016automated}.

Reducing the data volume requiring analysis by experts, or focusing their effort on the most pertinent data first, is necessary to improve the efficient of investigation. Quick et al. proposed an approach for digital forensic reduction through selective imaging~\cite{quick2016big}. The \textit{Select} process uses filters to display and select files to create a subset. These filters focus on artefacts from the file system; operating system (OS), software, Internet history, user created files, emails, documents, pictures, audio, video, etc. 
However, in many cases, e.g., child sexual exploitation material (CSEM) cases, the sheer amount of digital content is often still problematic after selective imaging multimedia file. This can result in significant psychological impact on the investigator; namely secondary traumatic stress disorder~\cite{sanchez2019practitioner}.

However, the data volumes can remain large after data reduction has been performed in some cases. Automated evidence analysis approaches are necessary for better categorisation of the volume of evidence. Beebe et al. use clustering for text searching results to rank evidence with their associated relevancy score, so as to improve the retrieval effectiveness~\cite{beebe2007digital}. The document's content can also be used for document clustering. Da et al. implemented a system whereby when a relevant file was found in a cluster, the investigator could prioritise the analysis of further files from the same cluster~\cite{da2012document}. Le et al. converted malware binary data to images for training deep learning models for malware classification~\cite{le2018deep}.

Timeline analysis is a process during the examination stage of an investigation that identifies the chronological events that have occurred on a device. Registry and log files, i.e., records of user actions, are used to build a timeline for further analysis. However, millions of low-level events can not be easily understood by investigators without knowing ground truth. Automated high-level digital event generation is one proposed solution~\cite{hargreaves2012automated}. 
File system traces also record an individual's actions on a device~\cite{casey2011digital}. For example, in a file download action, the date-time stamps of this file represent when the file was placed on the computer. File system metadata consists of a wealth of useful information for an investigation~\cite{rowe2011finding}. 

Data science is defined as: \textit{the ability to take data - to be able to understand it, to process it, to extract value from it, to visualise it, to communicate it}~\cite{loukides2011data}. Guarino et al.~\cite{guarino2013digital} identified that the big data challenge faced by digital forensics will lead to a convergence between data science and digital forensics, so as to resolve analysis of vast amount of data in actionable time. Sanchez et al.~\cite{sanchez2019practitioner} state that the sheer amount of digital content requiring analysis requires automatic forensic tools, artificial intelligence (AI) filtering, and safer presentation to practitioner.


Due to the aforementioned challenges faced, an automated approach for file artefact examination is required. In addition, the ability to quickly identify which file artefacts are likely to be most pertinent to the investigation at the earliest stage possible can greatly expedite the judicial process. This paper demonstrates an approach for file artefact prioritisation. 
The contribution of this work can be summarised as follows:

\begin{enumerate}
    \item The development of an approach for automated ranking of file artefacts by likely relevance, based on data reduction techniques.
    \item The development of a tool for automatically consuming information from generated timelines.
    \item An evaluation of the approach through experimentation with three emulated investigation scenarios.
\end{enumerate}

\section{Related Work}
\label{related}

\subsection{Digital Forensic Data Reduction}
\label{subsec:reduction}
The diversity of devices and sources of digital evidence results in corresponding diversity in digital forensic process models~\cite{du2017evaluation}. There is no single, universal process model suitable for all types of investigation. Reducing the volume of data for arduous, manual analysis will speed up the entire investigative workflow and can significantly aid in alleviating the digital forensic backlogs all too common in law enforcement agencies throughout the world~\cite{scanlon2016battling}.

Centralising digital forensic evidence processing enables investigators to take advantages of clustered performance and facilitates more efficient collaboration between the diverse roles in an investigation. A Digital Forensics as a Service system, \textit{HANSKEN}, has been developed and is currently in use for forensic investigation by the Netherlands Forensic Institute since December 2010~\cite{van2014digital,van2015digital}.

Data deduplication based on hash digestion comparison enables the reduction of unnecessary manually file examination. Hashing is a primary tool used in digital investigation~\cite{roussev2009hashing}. Hash-based techniques are used for a variety of purposes including finding known objects and finding similar objects, i.e., similarity hashing~\cite{lillis2018bloomfiltertrees}. The National Software Reference Library (NSRL\footnote{\url{www.nsrl.nist.gov}}) maintained by the US National Institute of Standards and Technology (NIST) contains a list of known hash values for most common OS and application packages. This list can be used to eliminate known, benign files.

A deduplicated digital forensic acquisition and analysis system capable of being integrated to a DFaaS system, e.g., such as \textit{HANSKEN}, was proposed in 2016~\cite{scanlon2016battling}. The framework eliminated the reacquisition of previously encountered and known files at the acquisition stage and enables the detection of illegal/pertinent file artefacts at the earliest stage of an investigation. Forensically sound disk image reconstruction from the deduplicated storage was proven to be possible using this system in 2018~\cite{du2018deduplicated}.

As the analysis focuses usually are different across various cases, another valid approach for data reduction is selective imaging of file artefacts depending on the investigative case type~\cite{quick2014data}. For example, in CSEM cases, Internet history logs, chat logs, Internet searches, images, movies files, calendars/notes; in narcotics cases, credit card information, electronic money transfers, financial records, fictitious identification, photographs of drugs and accomplices, unfilled prescriptions are more pertinent~\cite{alghafli2011guidelines}. 


\subsection{File System Metadata and Timeline Analysis}

In the analysis phase of a digital investigation, standard questions asked by the investigator include when, what, why, how? File system metadata records the most recent file actions, i.e., creation, access, and modification dates. Digital investigation looks to acquire information available on the system, from metadata and from timeline analysis to identify items of significant forensic value~\cite{buchholz2004role}.

File type allows investigators conduct data reduction. File system metadata including file size, file path, file name, etc., are usually used for filtering and indexing files in the examination stage of investigation. Directory metadata is used to find out the association between files, e.g., temporal association, spatial association, etc.~\cite{rowe2011finding}. 

OS and application log files also record the user's actions on a device. Data extracted from log files enable the generation of a timeline of the story on a device. Timeline visualisation can prove helpful for digital forensic investigation~\cite{olsson2009computer}. However, due to the typically large number of digital events extracted from a disk image, visualisation can often prove unhelpful in identifying pertinent events. As a result of each user action potentially generating several digital events on an abstract level, the number of timeline events is often too large for manual analysis. Millions of low level events are difficult to contextualise by investigators attempting to figure out the story on the device. Hargreaves et al.~\cite{hargreaves2012automated} outlined an approach for automatically generating higher level events, which greatly reduces their number -- making it significantly easier to be understood.

A combined timeline contains the digital events from several sources. \textit{log2timeline (plaso)}~\cite{gudhjonsson2010mastering} is a framework facilitating the generation of a ``super timeline'' including digital events from the file system, OS registry, logs, as well as application software logs. This contains information on both the device access level and the file system level. \textit{log2timeline} has been widely discussed in the field and forms the basis for significant further research. \textit{Timeline2GUI} was developed to analyse \texttt{*.csv} log files created by \textit{log2timeline}~\cite{debinski2019timeline2gui}. An abstraction based approach for timeline reconstruction was proposed in 2020, which is based on the timeline data provided by \textit{log2timeline}~\cite{bhandari2020abstraction}.

\subsection{Machine Learning in Digital Forensics}

Machine learning uses data features to build models to aid in specific tasks, e.g., a classification model for spam email recognition, a regression model for incoming email urgency assessment~\cite{flach2012machine}. Both classification and regression are supervised learning approaches, which requires the provision of labelled dataset and have been adopted to address problems in digital forensics.

Supervised machine learning in digital forensics investigation enables the leveraging of the results form the analysis phase. There are a couple of research approach outlined to assist the further investigation through training machine learning model using the previous investigation result. Marturanaet al.~\cite{marturana2013machine} presented an approach for digital device triage using machine learning. Devices are classified into criminal/noncriminal through machine features, which represent the user's habits, such as number of installed apps, max picture size, number of office/pdf files, number of compressed files, etc. Case studies on copyright infringement and CSEM exchange were also discussed. 

An approach for the automated determination of incriminating file artefacts is outlined by Du et al.~\cite{du2019methodology}. The file metadata is used as features for training classification models using known illegal and known benign files. The trained model is capable of recognising if previously unencountered file artefacts are likely to be pertinent to the investigation.

Machine learning as an automated solution for digital forensics shows significant promise to improve the efficiency of investigation. As stated by Flach~\citet{flach2012machine}: \textit{features are the workhorses of machine learning}. Leveraging the stored ``experience'' from the processing of previous investigations can facilitate the labelling of data for the training of automated classification models.

\subsection{Artefact Ranking/Prioritisation}
\label{subsec:2ranking}


For time-sensitive cases, pertinent information acquired from digital forensics has its greatest value at the earliest stage of the investigation. Triage is a process whereby devices and artefacts are ranked in terms of importance or priority~\cite{rogers2006computer}. Much work has been done in the area of digital forensic triage in an effort to improve the overall process~\cite{sanchez2019practitioner}. A digital forensic triage process model was proposed to use during the investigation by Rogers~\cite{rogers2006computer}. The importance of files varies in different types of case; CSEM, drug activity, financial crimes, etc. The approach for triage usually stems from practical experience. 

The triage process usually happens after a quick analysis of devices at the crime scene, then more in-depth analysis is performed in the digital forensic laboratory to identify more relevant evidence. When multiple devices are involved in an investigation, triage reduces the workload. Prioritisation of devices to be examined is defined as a sub-phase in the ``Behavioural Digital Forensics Model'' proposed in 2018~\cite{al2019behavioural}.

The larger the number of file artefacts encountered during an investigation, the more prolonged the examination process becomes. Image file examination is important for several cases types. 
In addition, keyword searching on file artefacts often results in a large number of results being returned. To analyse a large number of file artefacts in a limited time in an investigation, triage approaches enable to the prioritisation of effort. Search hit relevancy ranking algorithms was proposed by Beebe et al.~\cite{beebe2014ranking} for reducing the analytical burden of text string searching. A Support Vector Machine (SVM) model was trained for building the linear discriminant ranking function. The proposed feature list is based on past practice experience; 18 features were applied in the experiment for calculating the ranking score.

\section{Methodology}
\label{design}

The approach outlined in this paper aims to help in prioritising file artefacts requiring manual examination. It can be applied to an investigation after the data reduction phase. Data deduplication or hash database comparison steps can identify known benign and illegal file artefacts, and highlights previously unencountered files. Machine learning models can be trained against the known files and aid in the detection of the unknown files. The hypothesis is files with similar ``behaviour'' to illegal files are more relevant to the investigation, and should be recommended for further examination.

This approach consists of the steps listed below:

\begin{enumerate}
    \item Data deduplication and reduction, i.e., to get known files and unknown but interesting files.
    \item Disk image timeline generation, i.e., a ``Super Timeline'' generated by \textit{Plaso}.
    \item File artefact timeline generation.
    \item File artefact features extraction from the timeline.
    \item Model training using all known file artefacts.
    \item Relevancy score calculation on unknown , previously unencountered file artefacts using this model.
\end{enumerate}


\subsection{Overview of the Approach}


Comparing artefact hash values to a known database is a common approach to detect known illegal files during an investigation. The detection of known illegal/pertinent files can offer further insight than their mere presence for the further investigation on the device. The proposed approach in this paper takes advantage of these detected files to build a classification model, for identifying files that are similar to them and are likely more relevant to the investigation.

\begin{figure}[!h]
    \centering
    \includegraphics[width=0.5\textwidth]{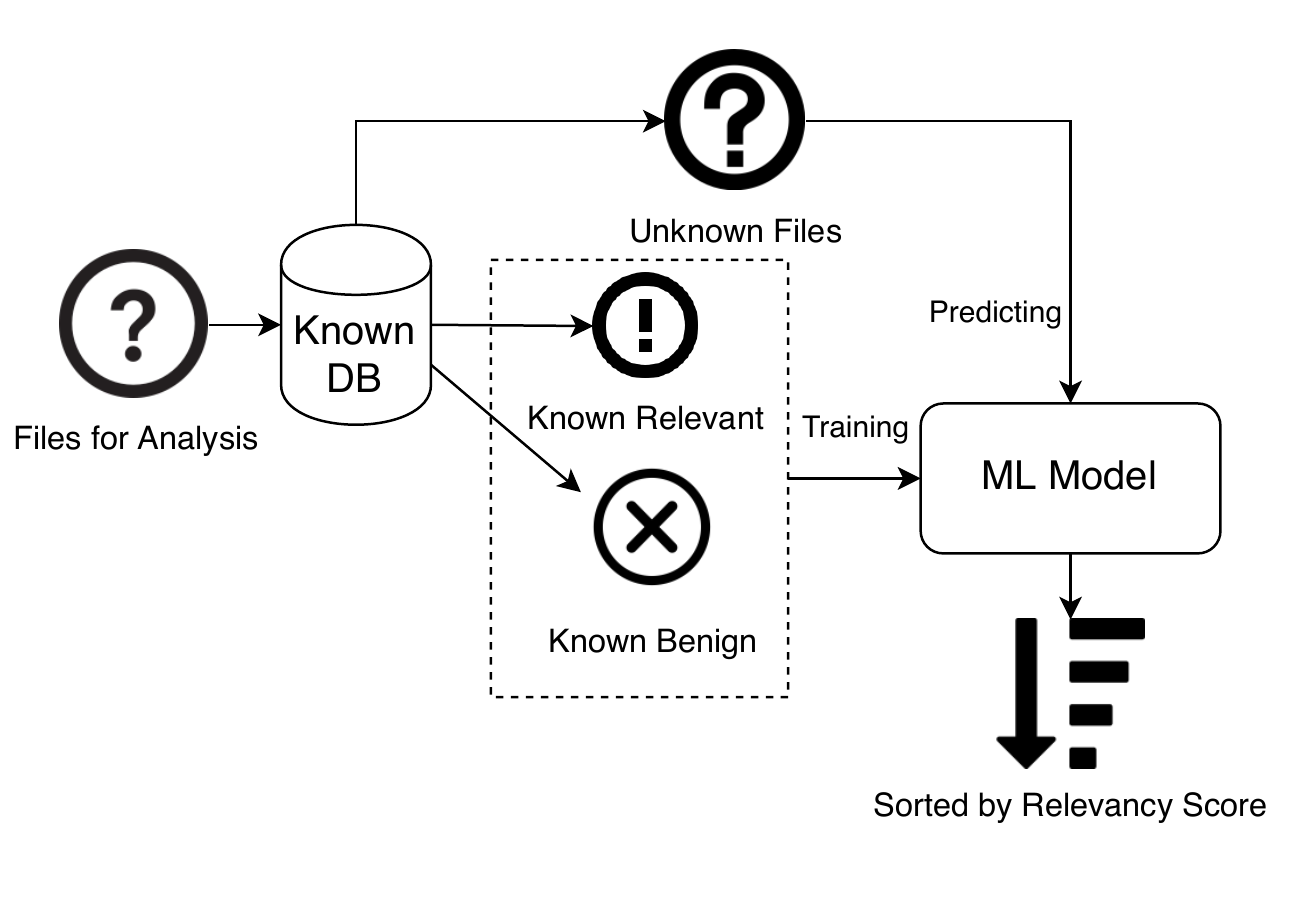}
    \caption{Overview of the Approach}
    \label{fig:approach}
\end{figure}

Figure~\ref{fig:approach} illustrates the approach, which takes advantage of database known files preserved from the analysis of previous investigations. The first step is to detect the known files by comparing the hashes on the target device to the known benign/illegal hash database. Secondly, using the digital events associated to the identified pertinent files to train a model for analysis of the unknown files. The trained machine learning model generates a relevancy score for each artefact, then they are by sorted by the score waiting to be analysed.

File artefacts that have associated digital behaviours more similar to the known illegal files are more relevant to the investigation. For detecting user behaviour for each file artefact, the device's timeline is filtered for those events pertaining to the artefact in question. The model is built using features extracted from each file artefact's timeline. Modification of content, metadata, access times, etc., can be obtained from this file artefact timeline. 

\subsection{Timeline Generation}

Existing forensic tools separately examine different type of artefacts -- such as tools focused on database forensics, email analysis, audio/video forensics, Internet browsing analysis, etc. Correlation of these disparate analysis results from different sources is a problem. ``Super Timeline'' generation creates a combined timeline using parsers to unify events from the diverse format log files. Analysis on this combined timeline can get general information instead of isolated details. Automating the analysis of this ``Super Timeline'' can aid in reducing the time wasted by manual analysis.

In this research, plugins for timeline analysis are developed for extracting the digital events related to each file. The generated file timeline consists of a complete story of available information for each file's creation, modification and access. The file timelines associated with the known file artefacts are used to build a machine learning model to predict if unknown file artefacts could be relevant to the investigation. 

\textbf{Disk image timeline}: A timeline of the disk image is generated using \textit{Plaso}. The timeline generated is stored in \texttt{csv} format for further analysis. Within this \texttt{csv} file, each row represents one particular event, including its source, type, description, etc.

\textbf{File artefact timeline}: One \texttt{csv} file is created for recording the digital events associated with each particular file. It is generated through searching for the file name (or historical names) in the timeline. The selected timeline features useful for the purposes of this work are:
\textit{inode, date, time, MACB, filename, type, source, sourcetype, datetime, and desc}. 


\subsection{File Artefact Feature Extraction}

\textit{Pandas}\footnote{https://pandas.pydata.org/} is the library used to to analyse the Plaso generated timeline, generating the file artefact timelines. Timeline-based features can be extracted by using the developed tool. These can be categorised as 3 types of features:

\begin{itemize}

\item Event Count: type of event, e.g., overall number of digital events, number of content modification events, number of metadata modification events, number of associated browsing history events, etc.

\item Datetime: When the event occurs matters, e.g., the file's creation time, last access time, etc. The conversion of the timestamp to a category value so it can be recognised by model is necessary. Features can be transformed to categories, e.g., date (month, workday, weekday), time (early morning, morning, afternoon, night, late night), etc.

\item Word count: A count of the occurrence of investigation specific keywords can also be used as features for model. As an example, for a drug related investigation, files may contain sales records, customer information, drug production instructions, or lists of precursor chemicals. These related words can be used as features. In addition, words discovered in known illegal files' timelines can be added to this keyword list.

\end{itemize}

\subsection{Feature Selection}

An abundance of features can be easily obtained from file artefact timelines. However, the number of features should fit the dataset to achieve optimal performance. Feature selection techniques can be applied to determine what features are best applied to model. Identifying the most influencing features can be used to improve the performance of a machine learning model. However, a balance must be struck -- having too few features in a model could lead to over-fitting. 

In this work, features stem from file timelines and file system metadata. The most popular events (and their corresponding source, type, etc.) can be used as features, but there are many non-pertinent events in a timeline, e.g., OS events. Feature selection helps to avoid missing useful features while also identifying the most significant features available. Which features and how many should be included? Generally larger datasets can handle more features. So when the dataset is small, fewer features can be beneficial to maintain a usable performance. 

\subsection{Relevancy Score}
\label{subsec:relevancy_score}

\begin{figure}[!ht]
    \centering
    \includegraphics[width=0.5\textwidth]{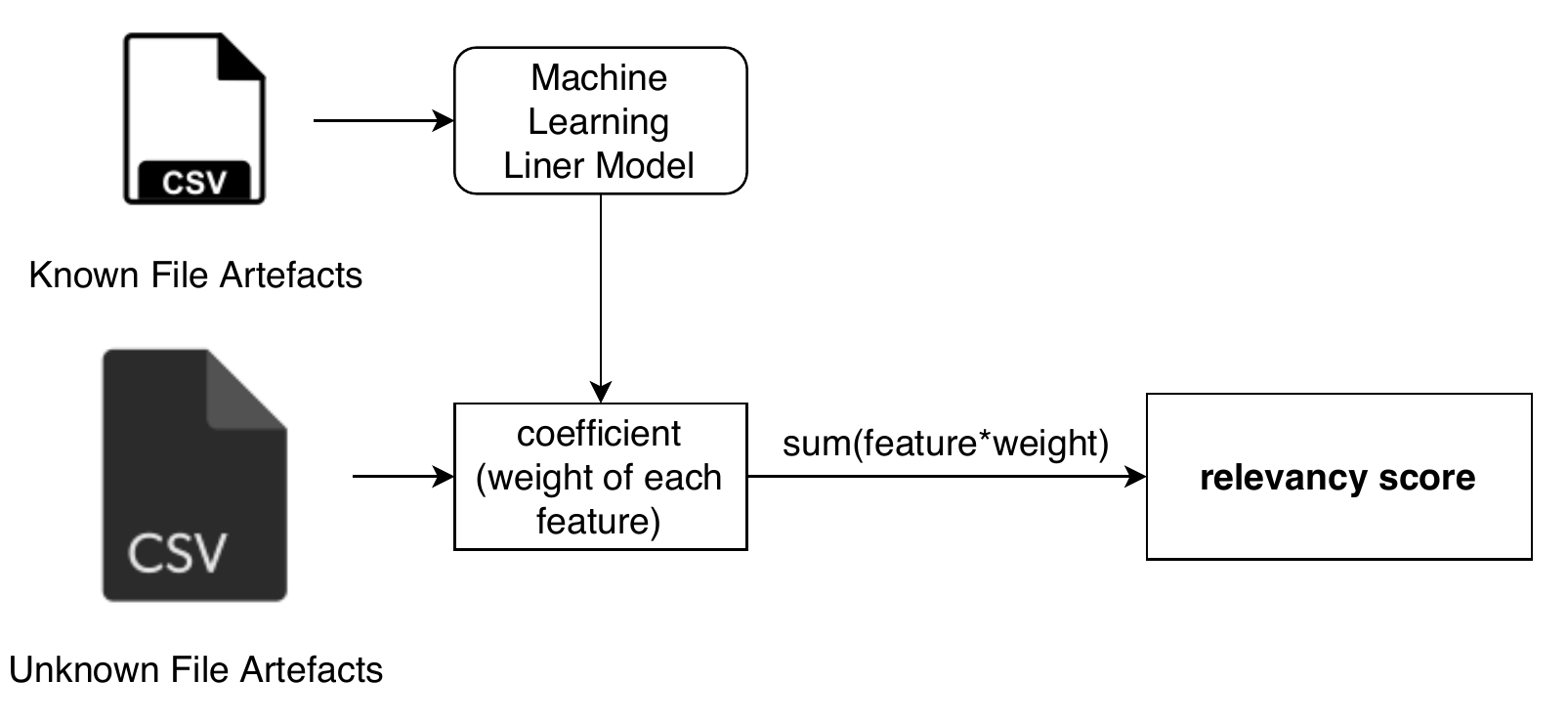}
    \caption{Relevancy Score Generation}
    \label{fig:score}
\end{figure}

Traditional triage or data reduction approaches builds filters based on investigative experience. For example, looking for document or an image in a financial crime, e.g., scanned documents, can cause an issue with the volume of results returned if merely filtering by file type. Specific keyword searching might only retrieve a very limited or empty result. In this research, a relevancy score is used to rank file artefacts when the number of files under examination is too large for manual trawling. This relevancy score is generated from a machine learning model trained by known file artefacts. It is a combination considering all given features, with more similar feature values resulting in a higher relevancy value, as can be seen in Figure~\ref{fig:score}). 

\textit{Scikit-learn}\footnote{https://scikit-learn.org/} offers the machine learning libraries used in this work. Coefficients can be obtained from linear modelling, i.e., Linear SVM, Logistic Regression, etc. Random Forest modelling also affords flexible determination of the significance of each feature. The script below shows how the coefficients of model are acquired:


\vbox{
\begin{code}
\SetupFloatingEnvironment{listing}{name=Snippet}
\begin{minted}[
bgcolor=LightGray,
fontsize=\footnotesize
]{python}
from sklearn import svm
svm = svm.SVC(kernel='linear')
svm.fit(features, labels)
svm.coef_
\end{minted}
\vspace*{-0.7cm}
\captionof{listing}{Obtaining the coefficients from the trained model}
\label{code:set_f_bios_time}
\end{code}
}

\section{Experimental Methodology}
\label{experiment}

In this section, the experimental design is outlined. Experimental data generation and processing is described in Section~\ref{subsec:data}; Example scenarios explaining how the proposed approach works and how it is tested presented in Section~\ref{subsec:case}.

\subsection{Experimental Data Generation}
\label{subsec:data}

Experimental data was generated manually specifically for the experiments outlined in this paper. Disk images from real cases or other shared digital corpus from second hand online purchases are not currently available for research due to data protection and ethical reasons. As the experiment requires analysis of its performance, emulated data is used that provides the following benefits: 1) files with illegal content are not needed, the proposed approach is using the files' associated digital events to determine if it is suspicious or not; and 2) generated data has a clearer and more detailed ground truth.

Disk images were generated as virtual machines (VMs). At first, actions were conducted generating files for investigation. Conducting similar operations on the same type file. This experiment aims to test the recognition of similar files through digital events -- therefore file metadata and content do not influence the experiment. Various files, with several file types, are randomly generated and downloaded onto the VM. Files with various user actions are emulated. General information for these files is listed in Table~\ref{tab:file_list}. These files in the VM are labelled as ``benign'', mixed with ``illegal'' file artefacts.

\begin{table}[!ht]
\centering
\begin{tabular}{c|c|c}
\hline
\textbf{File Type}& \textbf{User Actions} & \textbf{Number}  \\ \hline
pdf & creation (download from web) & 999 \\ \hline
txt & creation (notepad) & 100 \\ \hline
png & creation (download from web) & 100 \\ \hline
py & creation, access, run by python & 63 \\ \hline

\end{tabular}
\caption{``Benign'' File Information}
\label{tab:file_list}
\end{table}

The ``pertinent'' actions included emulated user activities for each of the three sample case scenarios (described in Section~\ref{subsec:case}). The actions defined as pertinent are those surrounding the activities with each scenario. For example, those associated with downloading CSEM (downloading research paper on the topic, picture download and photos sent/received using online chat tools); the execution of a hacking python script for cracking user's password; and creating fake invoices for a financial fraud investigation . The files related to these actions are labelled as ``pertinent''.

\begin{table}[!ht]
\centering
\begin{tabular}{c|c|c}
\hline
\textbf{File Type}& \textbf{User Actions} & \textbf{Number}  \\ \hline
txt & creation, access, edit & 6 \\ \hline
py & creation, unzip, access, move, copy & 6 \\ \hline
jpg & creation, access & 13 \\ \hline
png & creation (download from web), access & 4 \\ \hline
gif & creation (download from web), access & 1 \\ \hline
pdf & creation (download from web) & 1 \\ \hline

\end{tabular}
\caption{``Illegal'' File Information}
\label{tab:file_list}
\end{table}

\begin{figure}[!h]
    \includegraphics[width=0.5\textwidth]{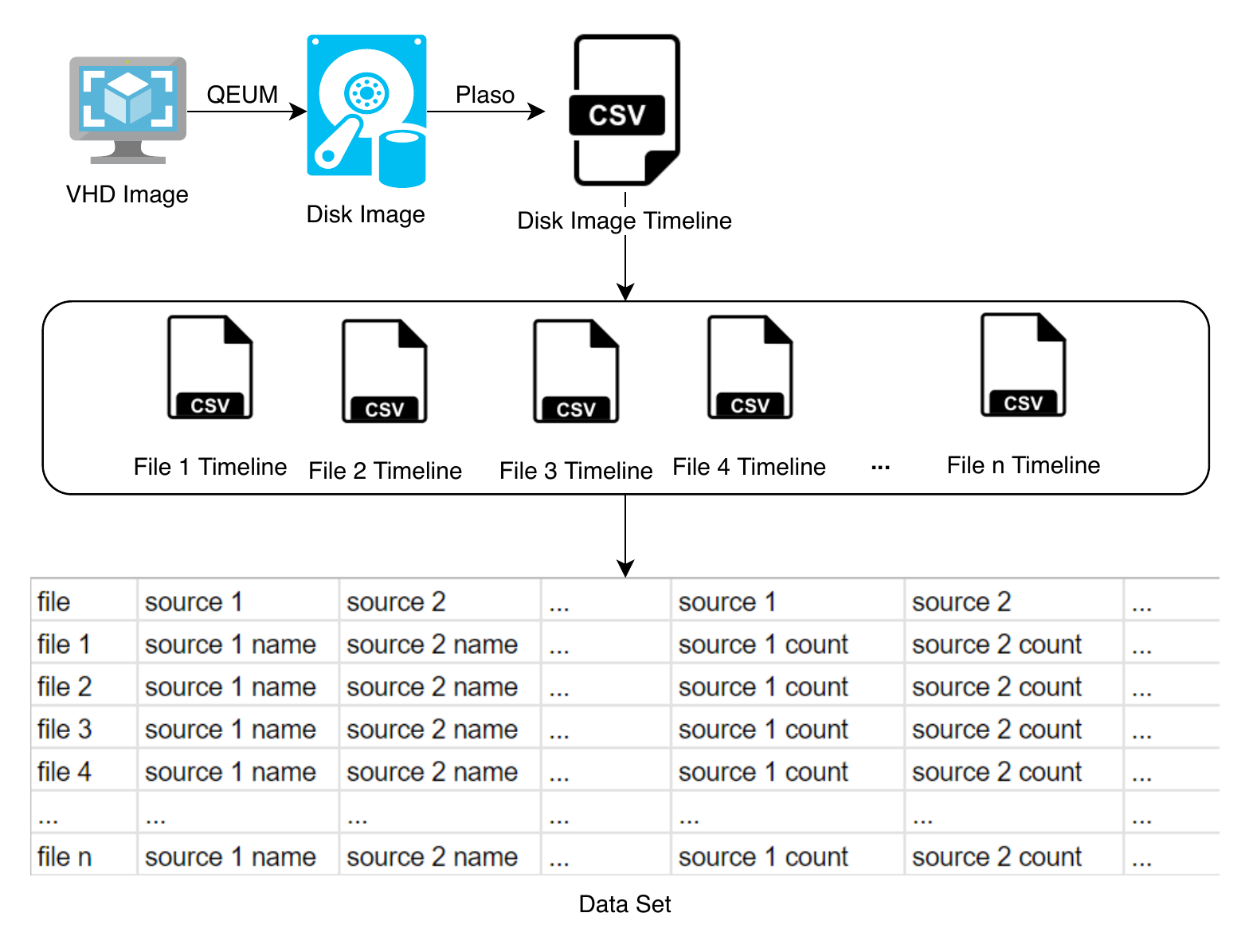}
    \caption{Data Set Generation Process}
    \label{fig:generation}
\end{figure}

Subsequently, the files' timelines were extracted. The process for this is shown in Figure~\ref{fig:generation}. The VMs' virtual hard disk (VHD) images were converted to raw format and had their associated timelines extracted using \textit{Plaso}. This timeline consists digital events associated with all of the file artefacts on the disk image. This is filtered for each file artefact to create the individual file artefact timelines. High frequency words and sources and the types of digital events are used as features for model training.

\subsection{Example Scenarios}
\label{subsec:case}
To demonstrate the viability of our approach, three sample scenarios are used (and will be referenced below):

\begin{enumerate}
    \item Possession of CSEM Investigation\\
            \textit{The suspect uses computer access chat room online related to child sexual exploitation material (CSEM). Videos and pictures are downloaded to local disk from installed browser. A computer belonging to a suspect was sized during a CSEM case investigation. Investigators use the known hash database filtering out the known illegal files; then a data reduction tool gets a set of user files that is most common to find pertinent files. These are chat log files, email files and picture files. With these picture files, some are detected as illegal from a known hash database. The investigator puts these file into a SVM model for training. In the end, other unknown files was put into the model, files are sorted by relevancy score for further analysis.}
    \item Hacking Case Investigation\\
            \textit{A computer was seized during a hacking case investigation. The suspect uses an email account. Keyword searching for ``username'' and ``password'' identifies several files. These are text files related to the use of password cracking scripts and scripts for hacking wireless networks. Investigators feed into a model to look for other similar files.}
    \item Financial Fraud Investigation\\
            \textit{The suspect creates a phishing site to con victims into supplying their email address and password and other personal information. The suspect uses their accounts to conduct fraud online. During an investigation of a financial fraud case, investigators are looking to find out potentially fraudulent financial instruments, invoices or other financial records. Searching the keyword ``invoice(s)'' in \texttt{pdf} and \texttt{doc} files from the raw disk image results in the discovery of some relevant files. Then investigators use the analysis result to build model to recognise similar files.}
\end{enumerate}

\section{Results}
\label{results}

In this section, 1) the experimental disk timeline acquired is presented showing the source of the the file timelines and what information is contained in a full disk image timeline; 2) the technique for generating these file timelines from the full disk timeline is presented showing the user action information; and 3) the case investigation process result is presented. 

\subsection{Disk Image Timeline Analysis}

This Section presents an overview of timeline generation. The full disk timeline reflects the usage of the seized machine, the number of digital events discovered in total, the number of files, the count of each digital events type, etc. \textit{psteal} is a tool in \textit{Plaso} for comprehensive disk image timeline generation and the command used is:

\texttt{psteal.py --source disk\_image\_name.dd -t l2tcsv -w timeline\_name.csv --partitions all}

From the generated timeline, basic information about the created disk image can be retrieved; in this scenario:

- Number of Events: 3,120,364

- Number of Files: 307,971

The timeline consists of all level of digital events. \textit{pandas} is used to further analyse the timeline. To acquire counts of unique values the method \texttt{count\_values()} is used. On a full disk image level, the source of digital events reveals information of the usage of device, such as \textit{Last Connection Time, Last Login Time, Last Password Reset}, etc. Digital events are related to file-system metadata information. In this scenario, the following were extracted:


\begin{table}[!h]
\small
\centering
\begin{tabular}{|c|c|}
\hline
\textbf{Event Type} & \textbf{Count} \\ \hline
Content Modification Time & 962,293 \\ \hline
Metadata Modification Time & 551,502 \\ \hline
\begin{tabular}[c]{@{}c@{}}Creation Time; Last Access Time; \\ Metadata Modification Time\end{tabular} & 343,906 \\ \hline
\begin{tabular}[c]{@{}c@{}}Content Modification Time; Creation Time; \\ Last Access Time; Metadata Modification Time\end{tabular} & 302,467 \\ \hline
Last Access Time & 283,980 \\ \hline
Creation Time & 235,871 \\ \hline
Content Modification Time; Creation Time & 212,687 \\ \hline
Creation Time; Last Access Time & 45,898 \\ \hline
\begin{tabular}[c]{@{}c@{}}Content Modification Time; Last Access Time; \\ Metadata Modification Time\end{tabular} & 35,818 \\ \hline
Last Access Time; Metadata Modification Time & 32,881 \\ \hline
\end{tabular}
\caption{File Artefacts Event - Common}
\label{file_events_common}
\end{table}

There are some types of events that can only occur to a specific type of file. For example, \textit{Previous Last Time Executed} could happen by a executable file, but not document or image file. Another example is a \textit{File Downloaded} event -- this occurs if a file is sourced from another machine through a network connection. These special events can be used as features pertaining to associated file artefacts, i.e., \textit{true} or \textit{false} as the feature value.

\begin{table}[!h]
\centering
\begin{tabular}{|c|c|}
\hline
\textbf{Event Type} & \textbf{Count} \\ \hline
Last Visited Time & 5,534 \\ \hline
Previous Last Time Executed & 1,107 \\ \hline
File Last Modification Time & 585 \\ \hline
Start Time & 410 \\ \hline
Last Time Executed & 401 \\ \hline
File Downloaded & 118 \\ \hline
Document Creation Time & 86 \\ \hline 
First Connection Time & 85 \\ \hline 
Document Last Save Time & 82 \\ \hline 
Content Deletion Time & 58 \\ \hline 
\end{tabular}
\caption{File Artefacts Event - Specific}
\label{file_events_specific}
\end{table}

    

\subsection{File Artefact Timeline Analysis}


This Section presents an example of file artefact timeline generation. It is the result of file digital events extracted from the full disk image timeline and outlines where the file features were extracted from.

The field \textit{filename} represents the source file of the digital event, instead of the file on which the event happens. For example, on a Windows machine using a NTFS file system, the file system metadata is from the \$MFT. The created file's filename is in the field \textit{desc}, i.e., description.

The file names are used to extract associated digital events. This action is conducted by use the file name as keyword to search each column of the file artefact attribute in the timeline. Action traces of file artefacts can be found and verified from the generated file timeline (consisting of various sources of digital events).

\subsection{Case Investigation and Relevancy Prioritisation}


This Section presents the results of the experimentation and investigative process conducted on each of the emulated case scenarios. 
For each case, a set of features are used considering the different investigation focuses. The features applied to model are determined by the detected pertinent files and what specific similarities/characteristics are looked for. Features extracted for building the model for each case are listed below:

\begin{enumerate}

    \item For the CSEM case scenario, the investigation focuses on images, videos, etc. The detected illegal files found have associated digital events from browsing activity. In addition, several file copying and moving actions for a number of the files were found in file timelines. For training the model to discover more files with a similar usage behaviour, the features used are: `chrome', `child', `png', `jpg', and `MFT'.

    \item In the hacking case scenario, python scripts for user password cracking and a couple of related text files were found. The python project was unzipped from a compressed file. Based on these details, the features used are: `hack', `python', `py', `txt', `zip', `unzip'.

    \item Investigation of the financial fraud scenario found emails that were sent with fake invoices (files in \texttt{pdf} format). The user had accessed the files close to time last use of the seized machine. The model building for further exploration uses features: `pdf', `invoice', `email', `fraud', `Last Access Time', `Creation Time'.
\end{enumerate}

The cases were tested on a dataset with 5.6\% of the files labelled as pertinent. Table~\ref{tab:result} shows that for each model, the recall metric obtained 75\% to 89\%, when only looking at the top 10\% of the resultant ranked result.

\begin{table}[!h]
\centering
\begin{tabular}{c|c|c|c}
\hline
 \textbf{No. Reviewed} & \textbf{Case 1} & \textbf{Case 2} & \textbf{Case 3} \\ \hline
  10\% & 0.75 & 0.82 & 0.89 \\ \hline
  20\% & 0.75 & 0.82 & 0.89 \\ \hline
  30\% & 0.79 & 0.82 & 0.89 \\ \hline
  50\% & 0.79 & 0.82 & 0.89 \\ \hline
100\% & 1.0 & 1.0 & 1.0 \\ \hline
\end{tabular}

\caption{Recall of each Model}

\label{tab:result}
\end{table}


\section{Discussion}
\label{discussion}


As seen in the previous Section, the results shows higher relevant file artefacts are effectively ranked to the top of the file list. These small data sets were generated for testing the variability of the proposed approach in simple use cases. Even though the amount of samples is small, this approach achieves an excellent performance in ranking the associated artefact. With a larger data set, abetter performance can be reasonably expected.

Feature selection is determined by the detected illegal files -- resulting in a model being created for relevancy score generation. The output from the process is a list of artefacts ranked by their relevancy scores -- indicating which to be expertly examined first. Of course, it is possible to miss some illegal artefacts solely relying on this approach. However, the pertinent files highlighted can be used to build a subsequent higher performance model.

\subsection{Benefits of this Approach}

This approach leverages the suspect device's ``super timeline'' that consists all levels of digital events, allowing comprehensive automated analysis on disk images. The approach outlined in this paper has the following potential benefits for digital forensic investigation:
\begin{itemize}
    \item \textbf{Automated analysis}: Automated device analysis on suspect devices performed immediately after acquisition can makes full use of the computation infrastructure available and can help prioritise the expert human investigator's focus during the analysis phase.
    
    \item \textbf{Data-driven approach}: Many existing tools can only obtain insights specific to a current case. For example, keyword search and filtering tools are limited to the current device under investigation and lose the insights learning for future investigations. A data-driven approach enables the detection of likely pertinent artefacts that are more difficult to be detected by traditional approaches by leveraging what has been processed before. Applying existing knowledge to explore new, previously unencountered data could prove fruitful in expediting the discovery process. 

    \item \textbf{Better performance as the known database grows}: The approach takes advantages of centralised evidence processing. The performance of this approach can be improved as the centralised dataset of processed cases gets bigger; a juxtaposition to the current digital forensic volume challenge. This is due to the bigger the known hash database gets, the higher the chance of detecting known pertinent file artefacts, the better the predictions can become. 

\end{itemize}

\subsection{Limitations of this Approach}


The objective of this work is to prioritise file artefacts and reduce the time needed for expert human file artefact examination. However, some limitations of the presented approach are observed:

\begin{itemize}
    \item \textit{Lack of known pertinent samples as input}: Known file artefacts are needed to train the machine learning models. The performance of the approach highly depends on the volume of previously analysed and categorised pertinent files.

    \item \textit{False positive and negative errors are possible}: Pertinent artefacts could be missed solely relying on this approach. However, as an evidence prioritisation/triage step, this approach can assist the investigation's focus. It is not intended as a substitute. In fact, both data reduction discussed in Section~\ref{subsec:reduction} and triage in Section~\ref{subsec:2ranking} are based on previous investigation experience. The purpose is to acquire actionable information at an earliest time possible.

\end{itemize}

Consequently, this approach should be used to assist investigation as a supplementary of the existing investigation tools. Manual analysis is still a necessity before and after using this tool, but it is envisioned that this approach can expedite the overall processes.

\section{Conclusion}
\label{conclusion}

This paper outlines an approach that prioritises file artefacts that are similar to previously analysed pertinent files. The automated process is assisted by developed feature extraction tools and machine learning models. The results show the advantages of the approach and indicates promise of expedited investigation. As a result, this approach would work best at an early stage in the examination to focus the investigation in promising directions.

\subsection{Future Work}
\label{future}

The approach in this paper present an automated analysis approach considering multiple sources of information. Additional sources of features could be included in this approach so as to expand its usability and accuracy. Further research is listed as follows:

\begin{itemize}
    \item \textit{Feature extraction from file content}: Further extensions of this approach will integrate the files' content as input features. For example, computer vision techniques can be applied on image and video file analysis and natural language processing techniques can be applied to document file analysis.
    
    \item \textit{Cross device analysis}: This can be conducted through analysis on a combined timeline from multiple devices. Seized devices and evidence sources from the same case or suspect can be joined together such as combining disk image artefacts with an email account, cloud service account, file transfer services, etc. 
    
\end{itemize}

\bibliographystyle{IEEEtran}
\bibliography{bibfile}

\end{document}